\documentclass[12pt,a4paper]{article}
\usepackage[utf8]{inputenc}
\usepackage[T1]{fontenc}
\usepackage{fix-cm}
\usepackage{amsmath,amssymb}
\usepackage{graphicx}
\usepackage{dcolumn}
\usepackage{bm}
\usepackage{hyperref}
\usepackage{doi}
\usepackage{geometry}
\usepackage{authblk}
\providecommand{\keywords}[1]{\noindent\textbf{Keywords:} #1}
\providecommand{\pacs}[1]{\noindent\textbf{PACS:} #1}

\newenvironment{acknowledgments}{\section*{Acknowledgments}}{}

\begin{document}

\title{Apokamp-type Gas Discharge Phenomenon:\\ Experimental and Theoretical Backgrounds}
\author[1]{Vasily~Yu.~Kozhevnikov}
\author[1]{Andrey~V.~Kozyrev}
\author[1]{Aleksandr~O.~Kokovin}
\author[1]{Aleksey~G.~Sitnikov}
\author[2]{Eduard~A.~Sosnin}
\author[2]{Victor~A.~Panarin}
\author[2]{Victor~S.~Skakun}
\author[2]{Victor~F.~Tarasenko}

\affil[1]{Laboratory of Theoretical Physics, Institute of High Current Electronics, Tomsk 634055, Russian Federation}
\affil[2]{Laboratory of Optical Radiations, Institute of High Current Electronics, Tomsk 634055, Russian Federation}

\date{\today}

\maketitle

\begin{abstract}
The apokamp discharge is an atmospheric-pressure plasma jet generated at the bending point of the discharge channel between high-voltage and floating-potential electrodes. The jet propagates perpendicularly to the interelectrode discharge channel without convection. In ambient air, the apokamp length reaches several centimeters, and the temperature of its tip is 100-250 °C. The typical apokamp propagation speed ranges from 100 to 220 km/s. The proposed time-dependent theoretical model explains the phenomenon. The discharge modeling confirms the tendencies observed in various experiments on the problem.
\end{abstract}

\maketitle

\section{\label{sec:level1}Introduction}

In 2016, an uncommon discharge type was first obtained experimentally in atmospheric-pressure air \cite{Sosnin_JETP}. It was a luminous structure formed near the bending point of the pulsed gas discharge channel, almost perpendicular to it. This type of discharge was called ``apokamp''\ (from Greek $\alpha \pi o$ - ``off''\ and $\kappa\alpha\mu\pi\eta$ - ``bend''{}).  

\begin{figure}[h]
\includegraphics[scale=0.25]{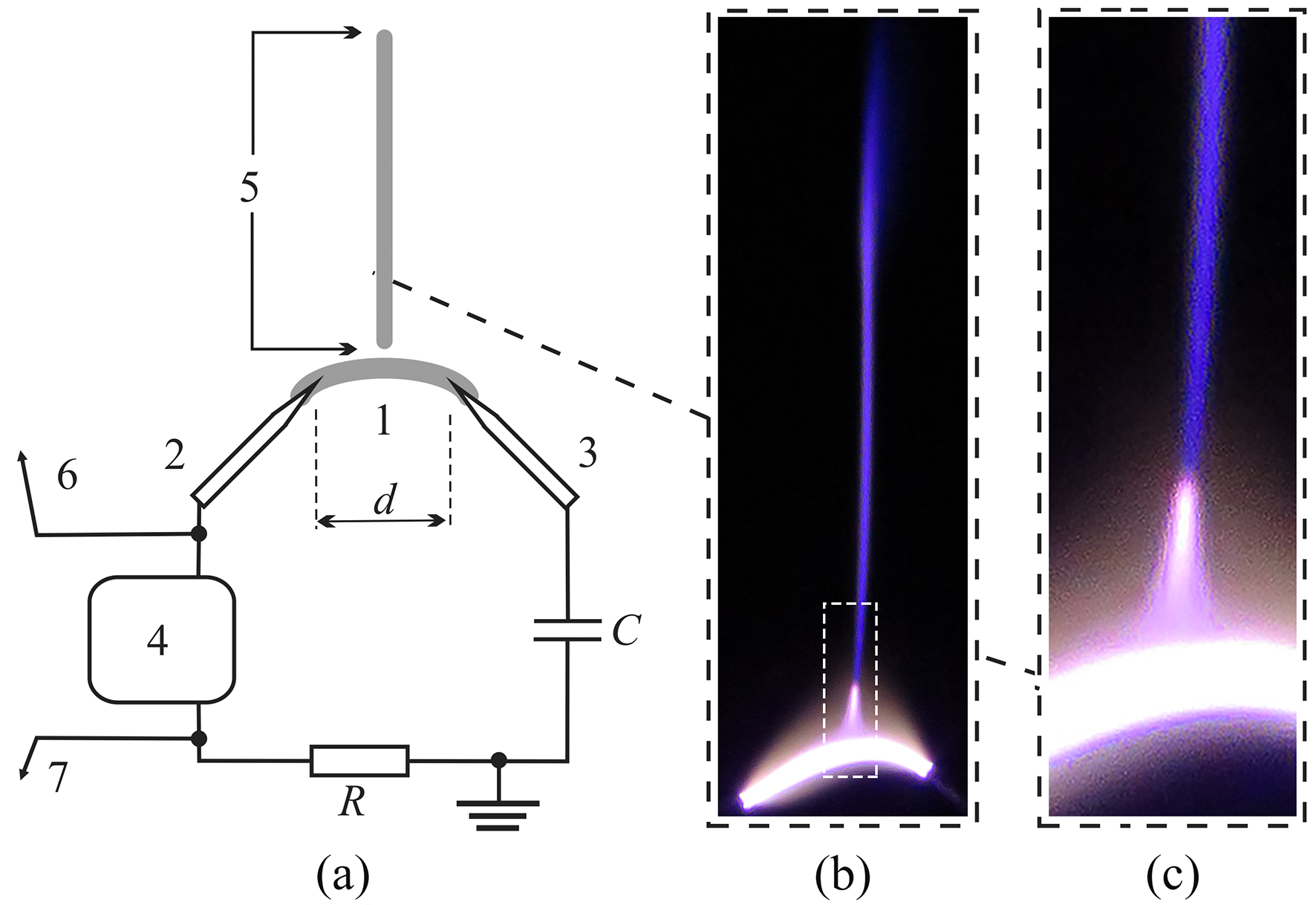}
\caption{\label{fig:experimental} Experimental setup for the apokamp obtaining under the atmospheric pressure (a), photographic image of the phenomenon (b) and zoomed area between discharge channel and plasma jet (c): 1 - discharge gap ($d = 8$~mm); 2, 3 - needle electrodes ($2$~mm diameter and $70~\mu$m tip curvature radius); 4 - pulsed voltage source; 5 - area of the apokamp formation; 6, 7 - terminals for voltage and current measurements, respectively ($C = 2$~pF, $R = 1$~Ohm).}
\end{figure}

Fig.~\ref{fig:experimental} shows the experimental setup that was used to obtain an apokamp discharge (a) and the external view of the jet (b, c). As it was shown to obtain an apokamp it is necessary to ignite a pulse discharge between electrodes 2 and 3, while these conditions have to be satisfied: 1) the applied voltage pulses at 2 is of positive polarity; 2) the discharge channel should have a natural or forced bend; 3) the electrode 3 and the discharge channel must be at under the potential of several kV w.r.t. the ``ground''. In Fig.~\ref{fig:experimental}, these conditions are provided by using a high-voltage generator producing positive polarity pulses with the voltage amplitude of $U_p \sim 11$~kV at a $f = 50.2$~kHz pulse repetition rate. The natural bending of the discharge channel was achieved by adjusting the angle between the electrodes to the range of $120 - 140^\circ$. As a result, an apokamp 5 in the form of a luminous plasma ``jet'' is formed near the channel bending point in atmospheric air. There is a dark intermediate region between the luminous region of the discharge channel and the beginning of the apokamp jet Fig.~\ref{fig:experimental}~(c). This photo was taken with a Canon PowerShot SX60 HS camera at a 1/8-second exposure and ISO 1500.

To estimate the plasma channel temperature, we introduced various substances into the channel and observed whether they were ignited or melted. In particular, the end of a nichrome wire with a length of $20$~mm and a diameter of $0.1$~mm was melted (the nichrome melting temperature is $ 1100-1400^\circ$~C) \cite{Sosnin_JETP}. So the lower channel temperature estimate was at least $1000^\circ$~C \cite{Sosnin_HT}.

Apokamp discharges were also experimentally obtained in various gas mixtures other than air (like $Ar-CO_2$, $Kr-N_2$, $Xe-Cl_2$, $Kr-Cl_2$) at pressures below atmospheric  \cite{Kusnezov}. The decrease in the molecular gas admixture deprives the jet of stability, leading to a transition from jet to volume discharge. Also, the important role of the electronegative gas admixture (mainly oxygen) in the formation and development of apokamp was noted.

Experimental papers \cite{Sosnin_EPJD, Tarasenko_PP} convincingly show that apokamp in air under normal conditions is an analog of a positive streamer whose tip is propagating into a non-ionized medium with characteristic velocities of $100-220$ km/s depending on the discharge operating voltage amplitude. Despite our extensive experimental background, the conditions for apokamp jet initiation remain unclear, as do the details of its subsequent development. In this paper, we aim to fill a long-standing gap in the theoretical treatment of this problem and to perform a deterministic plasma simulation to demonstrate the apokamp discharge phenomenon from first principles. 

\section{\label{sec:level2}Theoretical model} 
To understand quantitatively the formation of apokamp starting from the initial stage of the discharge ignition, we have used a ``two-moment'' non-stationary drift-diffusion macroscopic model \cite{Gogolides_1, Gogolides_2} implemented in the Plasma module of the COMSOL~Multiphysics~5.2a code. We provide a brief overview of the numerical computations, with details given in \cite{Kozhevnikov, UFN}.

For simplicity, we consider the computational domain to be two-dimensional rather than three-dimensional. It is represented by a rectangular cross-section area of $ 50\times70$~mm, including two blade-shaped electrodes with $0.1$~mm curvature radii that are located at an angle of 120~degrees to each other. The interelectrode distance between their tips is equal to $8$~mm. The floating potential (left) electrode is modeled as connected to ground via a $3.3$~pF capacitor. Another electrode is connected to the high-voltage source, which produces time-periodic $15$~kV-amplitude trapezoidal pulses with a $2.5~\mu$s duration. The electric circuit simulation is consistently implemented using the SPICE algorithm in the COMSOL~AC/DC~module.

Pure oxygen at $1$~atm was used as the gas medium in discharge simulations, as the most representative pure electronegative molecular gas. The existence of complete plasma-chemical cross-sections for oxygen, which involve numerous species and reactions, makes the simulation computationally expensive and unsuitable for 2D and 3D models \cite{Lee}. The scientific community has proposed different reduction techniques to simplify kinetic models of this type. Here, we use a minimal set of species and reactions for oxygen, including only those listed in Table~\ref{tab:table_plasmachem} of \cite{He}. These reactions are necessary to create an ion-ion discharge plasma with a small number of electrons, i.e., as is typical of electronegative gases.

Photoionization is another important reaction to consider when simulating plasma jets. In this model, the photon-molecule interaction $\hbar\omega + O_2\rightarrow e+O_2^{+}$ is implemented in terms of a ``differential'' representation given as a solution of a Helmholtz set of equations for the photoionization rate reaction \cite{Bourdon}. 

\begin{table*}[!htbp]
\caption{\label{tab:table_plasmachem} Reactions included in the model. For electron reactions, $T_e$ is the electron temperature in eV. }
\begin{tabular}{ccc}
Reaction & Threshold energy (eV) & Rate reaction ($\text{cm}^3/\text{s}$ or $\text{cm}^6/\text{s}$)\\ \hline \\
$e+O_2\rightarrow 2e+O_2^{+}$ & 12.06 & $9\cdot10^{-10}{T_e}^2\exp(-12.6/T_e)$ [$\text{cm}^3/\text{s}$]\\
$e+O_2+O_2\rightarrow O_2+O_2^{-}$ & 0.0 & $2\cdot10^{-29}(0.025/T_e)$ [$\text{cm}^6/\text{s}$]\\
$O_2^{+}+O_2^{-}+O_2\rightarrow 3O_2$ & - & $2\cdot10^{-25}$ [$\text{cm}^6/\text{s}$]\\
$e+O_2\rightarrow e+2O$ & 5.6 & $4.2\cdot10^{-9}\exp(-5.6/T_e)$ [$\text{cm}^3/\text{s}$]\\
\end{tabular}
\end{table*}

In this paper, the simulation of the apokamp initiation phenomenon is carried out for a single pulse from the voltage source, since the experiment indicates that apokamps are a sequence of fast-moving ``plasma bullets'', i.e., positive streamers moving in a pre-prepared plasma environment \cite{Sosnin_EPJD}. Namely, during the first $100-200$ voltage pulses, which follow at $\sim 25-50$~kHz repetition frequency, a spark ignition occurs between the electrodes multiple times, accompanied by the heating of near-electrode space to temperatures above $1100-1300^\circ$~C, the appearance of a surrounding luminous halo, and bending of the main discharge channel without an apokamp growth \cite{Tarasenko_PP}. 

To simulate starting conditions for the initial pre-ionization and to fix the neutral gas pre-heating, the symmetric temperature and the number density of the quasi-neutral ion-ion plasma in the form of ``Gaussian spots'' have been set. The background temperature is equal to $T_0=300$~K and background plasma number density is ${n_0\sim 10^3-10^4}\text{ cm}^{-3}$, maximal temperature of main discharge channel $T_{max}=1100-1300^\circ$~C is obtained from experimental data, while the maximal number density $n_{max}\sim 10^{12}\text{ cm}^{-3}$ corresponds to typical plasma density values for self-sustained atmospheric discharges. Note that the proposed model does not explicitly account for gas heating; instead, it uses a predefined constant-temperature field. The position of the plasma initial ``spot'' is chosen directly in the middle of the interelectrode gap. The initial plasma distribution is an elliptical spot stretched horizontally, taken as the closest shape to that of a real plasma channel produced by many voltage-pulse repetitions. The same has been done for the heating profile, i.e., located slightly above the distance between the electrode tips.

\begin{figure}[ht]
\includegraphics[scale=0.5]{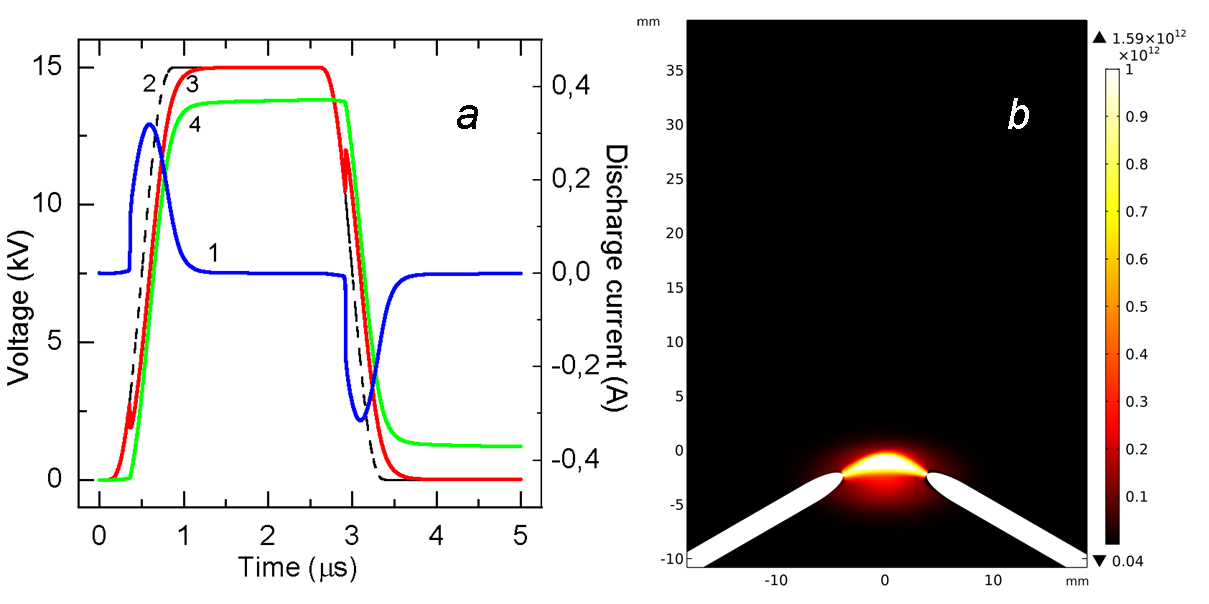}
\caption{\label{fig:noapokamp} Time-resolved profiles of the current and voltage in a pulsed-periodic gas discharge (\textit{a}) 1~–~discharge current, 2~–~source voltage, 3~and~4~-~ electricic potentials at the respective electrodes) and (\textit{b}) the spatial distribution of the ion–ion plasma. }
\end{figure}

Fig.~\ref{fig:noapokamp} shows the simulation results in the framework of the theoretical model proposed above. The discharge time profiles in Fig.~\ref{fig:noapokamp}~(\textit{a}) are distinctive for the current flow in a two-electrode system with a floating potential electrode. When voltage is applied to the high-voltage electrode, discharge ignition occurs, accompanied by simultaneous charging of the capacitance connected to the second electrode. After the capacitance is charged to the first electrode's potential, the discharge quenches. It ignites again only after a significant decrease in the source voltage, so the gap voltage drop reaches the breakdown value.

In Fig.~\ref{fig:noapokamp}, the discharge channel has a typical bend, and a plasma halo surrounds the near-electrode space according to the experimental pictures in Fig.~\ref{fig:experimental}~(c).  But there is no formation of a plasma jet from the bending area of the discharge channel. The plasma distribution remains unchanged even as we increase the number of voltage-pulse repetitions to tens or hundreds. 

\begin{figure}[ht]
\includegraphics[scale=0.24]{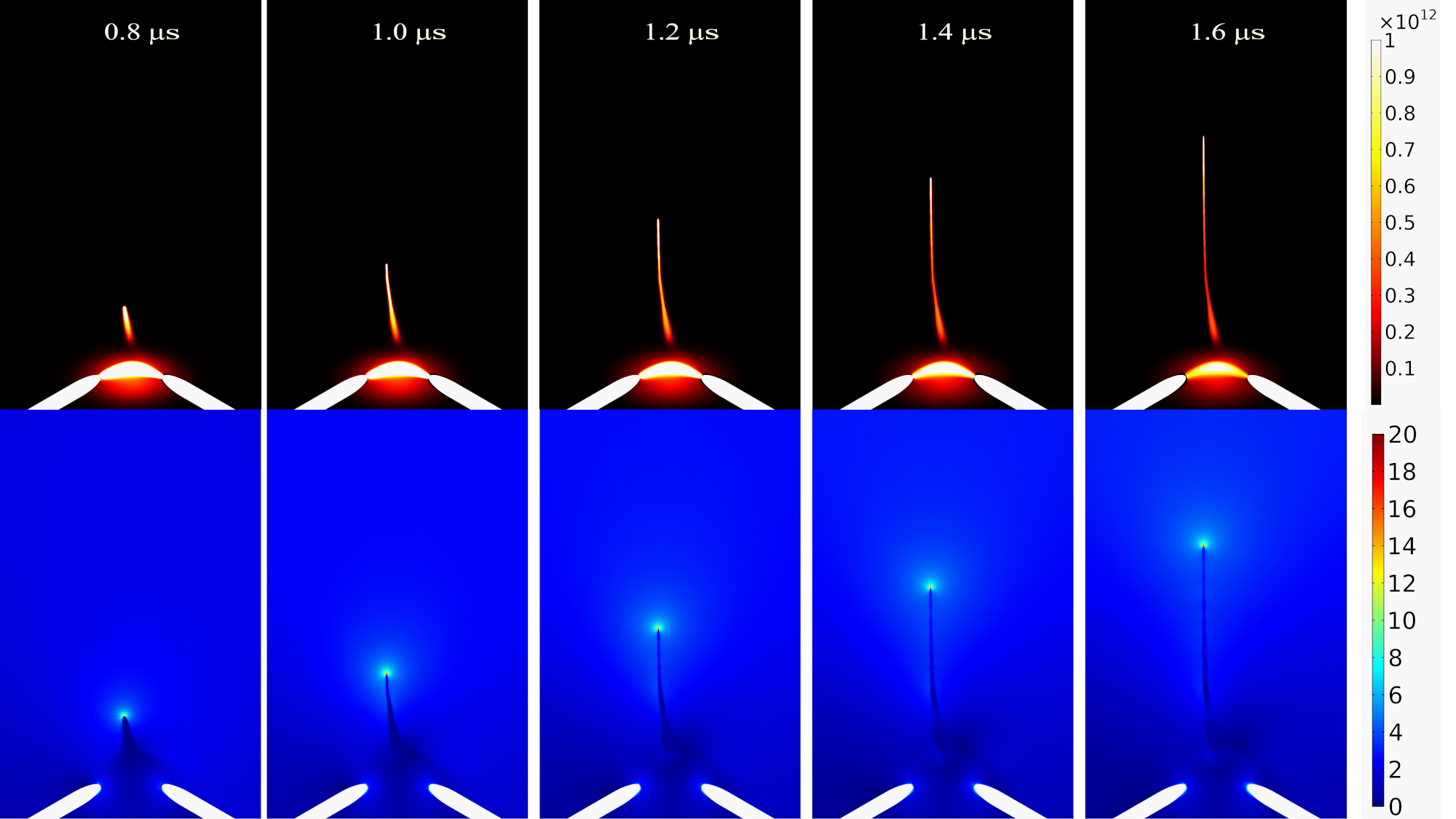}
\caption{\label{fig:apokamp} The sequences of frames illustrating the phenomenon of apokamp growth during a single voltage pulse: distribution of ion-ion plasma (top row, in $\text{cm}^{-3}$) and the absolute value of electric field (lower row, in kV/cm).}
\end{figure}

Since the experiments convincingly show that the phenomenon's nature is not related to gas convection, it follows that apokamp formation requires an external electric field directed perpendicular to the main discharge channel. To determine the role of a weak external transverse electric field, the electric potential of the upper boundary of the computational domain was set to zero (grounded). As the discharge channel between the metal electrodes was located about $60$~mm from the upper boundary of the computational domain, the difference in electric potentials was approximately ten times smaller than the breakdown voltage for this gap. Nevertheless, the presence of a third electrode, placed far from the discharge, led to the appearance of a weak transverse electric field and the development of a plasma jet channel.

Fig.~\ref{fig:apokamp} shows the dynamics of the main discharge ignition followed by the plasma halo formation and growth of the apokamp jet towards the third (far) electrode. For a more detailed description of this phenomenon, both the distribution of plasma components and the absolute value of the electric field are given in Fig.~\ref{fig:apokamp} for the selected time points.

The growth of apokamp (Fig.~\ref{fig:apokamp}) occurs during a period of time less than one voltage pulse duration. The ``appendix'', i.e., the base of apokamp, starts at the $500$~ns time point after the main discharge ignition. There is a less dense ($\sim 5\cdot 10^{11} \text{ cm}^{-3}$) region between the dense plasma region of the main discharge channel and the nascent apokamp appendix. The appendix appears in the region where the electric field reaches its relative maximum, between the plasma halo boundary and the surrounding non-ionized gas. Further apokamp development is driven by the intensive oxygen photoionization typical of positive streamer dynamics. Both the appendix and the whole apokamp jet are made of dense oxygen plasma ($\sim 10^{13} \text{ cm}^{-3}$), mainly constituted by positive and negative ions. 

In the second row of Fig.~\ref{fig:apokamp}, the distribution of electric field (absolute value) is depicted. As can be seen, the tip of the apokamp channel has an electric field of $\sim 20$~kV/cm, i.e., sufficient to effectively ionize the surrounding gas and propagate into the non-ionized medium.

\begin{figure}[ht]
\includegraphics[scale=0.225]{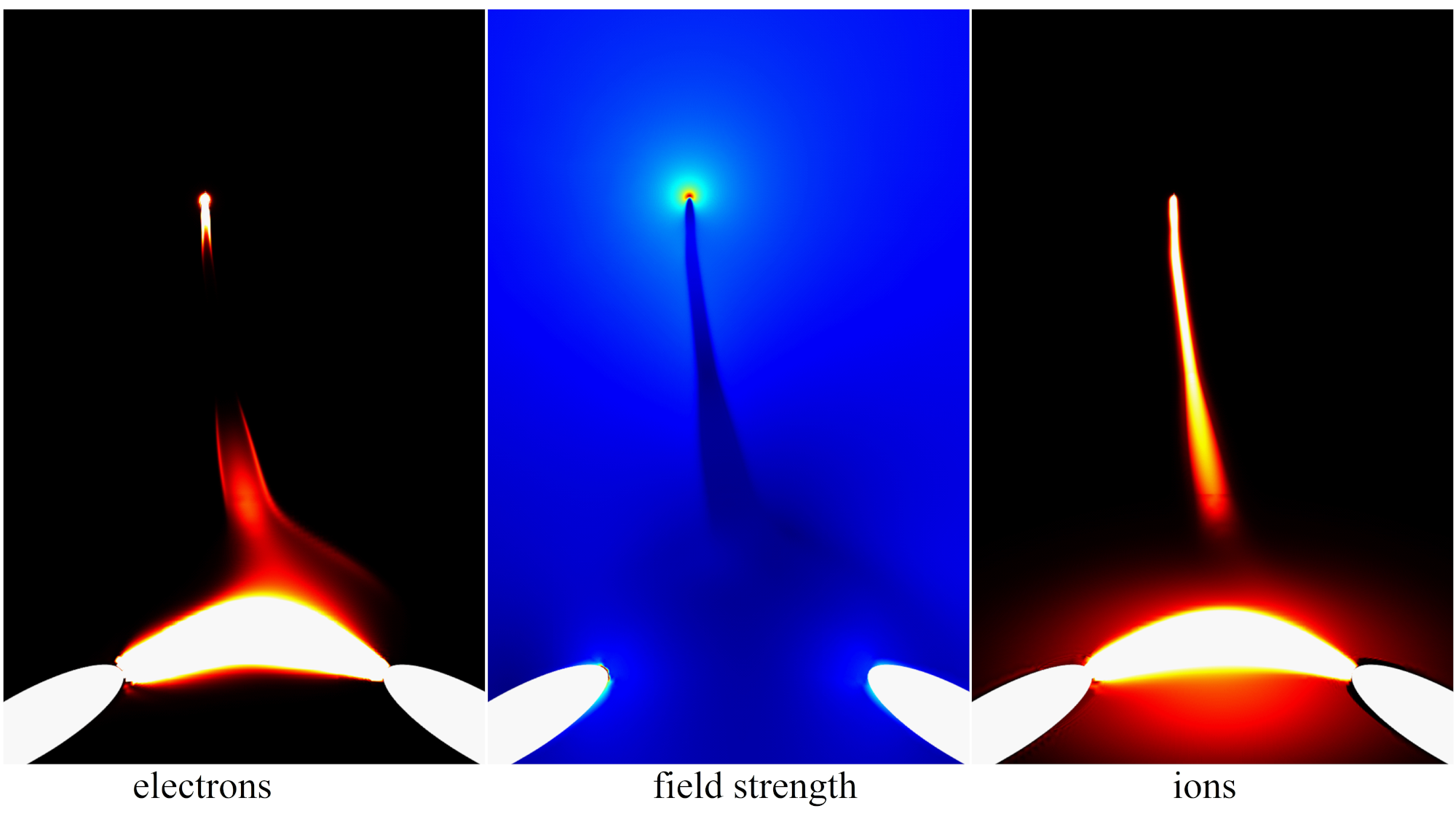}
\caption{\label{fig:bullets} Detailed distribution of electrons, electric field, and ion-ion plasma for a $1~\mu$s time point simulation of an apokamp plasma channel formation. }
\end{figure}

Fig.~\ref{fig:bullets} illustrates the evolution of electron number density as well as ion plasma and the electric field at $1~\mu$s. It can be seen that the electron density distribution reaches its maximum only at the tip of the growing apokamp channel, which propagates at approximately $30$ km/s. The increased electron density at the tip facilitates excitation of metastable atomic levels, leading to short-term localized luminescence in the corresponding region. Visually, on high-speed photography, such regions \cite{Sosnin_EPJD} look like plasma clots or plasma ``bullets'' that are experimentally observed in this phenomenon at atmospheric pressure.

The rapid electron attachment inside the growing streamer channel maintains a slow-decaying ion-ion (electronegative) plasma, providing a thin apokamp channel. It is known that the local amplification coefficient of the electric field at the top of a long conductive rod strongly depends on the ratio of the length of the tip, i.e., $l$, to the radius of its cross section, $r<<l$, namely, $K\sim (l/r)$ \cite{Chatterton_1, Chatterton_2}. Just the small transverse size of the channel, $r$, provides sufficient weak external-field amplification to the breakdown field at the top of the growing apokamp channel.

\section{\label{sec:level3}Conclusions.}

We have presented a first-of-its-kind simulation of the apokamp discharge phenomenon, previously discovered in experiments. The simulation has demonstrated that apokamp is a phenomenon that arises in a weak external electric field and originates from the main channel of the gas discharge with a floating-potential electrode. If these conditions are met, the resulting apokamp jet exhibits a stable, electronegative plasma structure and propagates into non-ionized regions at high speed. This creates significant electric fields on the tip. The simulation convincingly explains the details of apokamp jet formation observed in experiments, such as the propagation of ``plasma bullets'' in the apokamp channel, as recorded by high-speed imaging. Also, the simulation revealed a nontrivial spatial structure of the apokamp jet, which is established at earlier stages of jet formation.

Based on the electrical and optical measurements of the apokamp discharge parameters, as well as from the theoretical simulation results, we can draw the following brief conclusions:
\begin{enumerate}
\item The apokamp is a narrow streamer channel that grows at characteristic velocities ranging from $20$ to $200$~km/s (depending on the applied voltage, pressure, and neutral gas type) from the bending point of a discharge in pulse-repetition mode. The bending region provides a local electric field enhancement setting the initial anisotropy of the starting streamer;

\item The necessary conditions for an extended streamer channel growing are created by a relatively weak ($\sim 2$~kV/cm, an order lower than static breakdown values) macroscopic electric field maintained between the high-voltage discharge channel and the far zero-potential surrounding space. This low average field enhances at the growing streamer head to near-breakdown values ($\sim 20$~kV/cm, which is about $70$~\% of the static breakdown threshold), thereby ensuring rapid growth due to local gas ionization.

\item Most likely, in non-attaching gases, the apokamp is either not observed at all or is not so clearly marked due to the rapid expansion of the electron-ion plasma in the transverse direction. A smaller field enhancement coefficient cannot ensure a high gas ionization rate at the streamer tip.

\item The pulse-periodic discharge power supply mode reproduces the streamer channel (apokamp) during every single pulse, but the preceding ion-ion plasma core provides reproduction of the channel shape and direction from pulse to pulse. Therefore, the pulse-periodic mode discharge looks like a continuous plasma jet with a characteristic shape tied to the bend of the main discharge channel.
\end{enumerate}

\begin{acknowledgments}
This research has been supported by the program of State assignment of the ISE~SB~RAS, project No.~FWRM-2026-0008, FWRM-2026-0009.
\end{acknowledgments}

\bibliographystyle{unsrt}
\bibliography{apokamp}

\end{document}